# Nature of magnetic exchange interactions in kagome antiferromagnets FeGe and FeSn

Yitao Zheng[a], Yan Zhu[b,*], Jun Hu[a,†]

[a] Institute of High Pressure Physics, School of Physical Science and Technology, Ningbo University, Ningbo 315211, China

[b] College of Science, Nanjing University of Aeronautics and Astronautics, Nanjing 210016, China

Magnetic exchange interactions (MEIs) in kagome magnets exhibit rich features due to the interplay of charge, spin, orbital and lattice degrees of freedom, giving rise to a variety of exotic quantum states. Through first-principles calculations, we systematically investigate the MEIs in kagome antiferromagnets FeGe and FeSn. While the antiferromagnetic order originates from the interlayer coupling between neighboring kagome layers, Fe atoms within each kagome layer couple ferromagnetically, driven by the competition between ferromagnetically favorable direct MEIs and antiferromagnetically favorable Ruderman-Kittel-Kasuya-Yosida (RKKY) interactions. The stronger direct MEIs but weaker RKKY interactions in FeGe result in a substantially higher Néel temperature with respect to FeSn. Interestingly, the nearest neighboring exchange energy in both materials approximately linearly depends on the Fe-Fe bond length, so that moderate compressive strain can significantly enhance their Néel temperatures.

* yzhu@nuaa.edu.cn

† hujun2@nbu.edu.cn

## I. Introduction

A kagome lattice is an extended two-dimensional (2D) honeycomb network in which atoms occupy edge centers of hexagons rather than their vertices, forming a pattern of corner-sharing triangles interlaced with hexagons, as shown in Fig. 1(a). First introduced by Syôzi in 1951 in the context of solveing the 2D Ising model, the kagome lattice was later observed in three-dimensional (3D) material—jarosite compounds $RFe_3(OH)_6(SO_4)_2$(R=NH$_4$, Na or K)—in 1968.[1,2,3] Owing to its distinctive atomic symmetry, the kagome lattice provides an ideal platform for the interplay between charge, spin, orbital and lattice degrees of freedom, inducing a variety of exotic quantum states. These states include strongly correlated electronic phases[4,5], diverse magnetic orders[6], and unconventional superconductivity[7,8,9,10]. Additionally, the kagome lattice can host a wide range of topological phases[11,12], such as fractional quantum Hall states[13], quantum anomalous Hall effect[14,15], and Dirac or Weyl fermions[16,17,18]. Such phenomena hold great promise for next-generation electronic and spintronic applications, and have driven intense research interest in kagome-based materials[11,19,20].

Among various kagome materials, Co-Sn-type intermetallic compounds, including CoSn, FeGe, and FeSn, have drawn particular attention for their unusual electronic properties[21]. This structural family consists of alternately stacked kagome and honeycomb layers. The kagome layers host an additional Ge or Sn atom at each hexagonal center (notated as $Ge_k$ or $Sn_k$), while the honeycomb layers are composed of Ge or Sn atoms (notated as $Sn_h$ or $Ge_h$), as shown in Fig. 1(c) and 1(d). It has been revealed that both FeGe and FeSn are metallic A-type antiferromagnets, with Néel temperature ($T_N$) of 410 K and 368 K, respectively[22,23,24,25]. In this antiferromagnetic (AFM) order, the spins of Fe atoms align ferromagnetically within each kagome layer but antiferromagnetically between adjacent kagome layers[5,26]. Intriguingly, FeGe exhibits an additional charge density wave (CDW) order below 100 K, characterized by Ge-dimerization between Ge atoms in adjacent kagome layers[28,29]. This CDW order originates from Fermi surface nesting, and leads to local gap opening at the van Hove singularity near the Fermi level ($E_F$)[30,31], which in turn enhances the magnetic moment of Fe atoms. However, a comprehensive theoretical study on the magnetic exchange interactions (MEIs) and the intertwining between the MEIs and the CDW order in these materials remains absent.

In this paper, we carried out first-principles calculations to systematically investigate magnetic properties of FeGe and Fesn, including MEIs, the influence of the CDW phase, and the effect of biaxial strain. We found that the intra-layer FM coupling is dominated by nearest neighboring MEIs, arising from the competition between ferromagnetically favorable direct MEIs and antiferromagnetically preferable Ruderman-Kittel-Kasuya-Yosida (RKKY) interactions[32,33,34]. The significantly stronger RKKY interactions in FeSn result in remarkable reduction of $T_N$ compared with FeGe. Moreover, the CDW phase of FeGe further enhances its magnetic state at low temperatures. Through applying biaxial strain, we revealed a unified dependence of the local spin moment of Fe atoms in both materials on the Fe-Fe bond length within kagome layers. Furthermore, the nearest neighboring exchange energy is approximately linearly dependent on the Fe-Fe bond length. Interestingly, biaxial strain remarkably boosts $T_N$ of both FeGe and FeSn, manifesting that it is an effective approach to engineer magnetic orders in kagome magnets.

## II. Computational details

First-principles calculations were performed within the density functional theory (DFT) framework using the Vienna *ab* initio Simulation Package (VASP)[35,36]. The interaction between valence electrons and ionic cores was described by the projector augmented wave (PAW) method[37,38]. The Bloch wavefunctions were expanded using a plane-wave basis set with a kinetic energy cutoff of 500 eV. The exchange-correlation potential was treated at the generalized gradient approximation (GGA) level with the Perdew–Burke–Ernzerhof (PBE) functional[39]. Convergence criteria were set at 0.01 eV/Å for forces and $10^{-5}$ eV for total energies. The Brillouin zone (BZ) was sampled by a fine Γ-centered 15×15×6 Monkhorst−Pack mesh. The exchange energies were calculated using the TB2J package[40], which is based on the maximally localized Wannier functions as implemented in the Wannier90 code[41].

## III. Results and discussion

We first optimized the atomic structures of FeGe and FeSn, obtaining lattice constants of *a* = 4.96 Å and *c* = 8.11 Å for FeGe, and *a* = 5.29 Å and *c* = 8.93 Å for FeSn, in good agreement with the experimental measurements[23,26,28]. Although the AFM order in FeGe and FeSn is closely associated with electron correlations, traditional DFT calculations without the Hubbard U correction have successfully reproduced the Fe magnetic moments — approximately 1.5 $\mu_B$ for FeGe and 2.0 $\mu_B$ for

FeSn[23,27,30]. Therefore, we adopted the standard DFT approach to investigate the electronic and magnetic properties of FeGe and FeSn.

Figure 2(a) displays the band structure of FeGe, revealing its metallic character with a series of flat bands in the relevant energy range. Additionally, three Dirac points are located below $E_F$ at the K and H points of the BZ, as labeled in Fig. 1(b). Comparing the projected density of states (PDOS) shown in Fig. 2(b-d), these flat bands originate from Fe-3d orbitals. The PDOS of the $Ge_k$ atom (i.e. in the kagome layer) exhibits clear spin polarization, while that of the $Ge_h$ atom (i.e. in the honeycomb layer) remains non-spin-polarized, as depicted in Fig. 2(c) and 2(d). Both the band structure and PDOS of FeSn exhibit similar features to those of FeGe.

To examine the MEIs, we calculated the exchange energy J between two Fe atoms as a function of their separation vector $\vec{r}$, denoted as $J(\vec{r})$. Owing to the crystal symmetry, the Fe-Fe distances are actually a set of discrete values. For short Fe-Fe separations, $J(\vec{r})$ can be categorized as intra-layer ($J_i$) and inter-layer ($J_{ci}$) interactions, representing the exchange energy between the *i*-th nearest Fe neighbors in the same kagome layer and in different kagome layers, respectively, as depicted in Fig. 1(c) and 1(d). Furthermore, Fig. 1(c) shows that Fe atoms are arranged differently along different directions. Accordingly, $J(\vec{r})$ can be approximated as the sum of $J_{<100>}(r)$, $J_{<110>}(r)$, and $J_{<120>}(r)$ for three intra-layer low-index directions, and $J_{<001>}(r)$ for the inter-layer directions, as labeled in Fig. 1(c). Here, these directions are defined relative to the Fe network rather than conventional crystallographic directions. For instance, <100> refers to Fe chains running along the edges of Fe triangles, whereas <110> corresponds to Fe-Ge chains passing through hexagonal centers.

The calculated $J(\vec{r})$ values for both FeGe and FeSn along these directions are plotted in Fig. 3. In all cases, $J(\vec{r})$ decreases rapidly as the Fe-Fe distance increases. The intra-layer $J_i$ and inter-layer $J_{ci}$ values highlighted in Fig. 1(c) and 1(d) dominate over the rest points which have longer Fe-Fe distance. As listed in Table I, $J_1$ is significantly larger than all others — 8.99 meV for FeGe and 7.92 meV for FeSn. The positive sign of $J_1$ indicates that the nearest Fe neighbors favor ferromagnetic (FM) coupling. Although some $J(\vec{r})$ values such as $J_2$ and $J_3$ are negative, implying AFM coupling between the corresponding Fe-Fe pairs, their magnitudes are much smaller than that of $J_1$. Consequently, the overall intra-layer magnetic configuration favors an FM state. On the contrary, $J_{c1}$, representing the MEIs between the nearest inter-layer Fe-Fe pairs, is -1.25 meV for FeGe and -1.12 meV for FeSn, resulting in AFM coupling between adjacent kagome layers. Furthermore, the magnitudes of both $J_1$ and $J_{c1}$ in FeGe are larger than those in FeSn, resulting in much higher $T_N$ in FeGe.

Note that $J_1$ usually represents the direct MEI, whereas all other $J(\vec{r})$ points arise from indirect MEIs, such as superexchange interaction. Interestingly, $J(\vec{r})$ exhibits an apparent oscillatory behavior beyond distances exceeding the lattice constant, suggesting that long-range MEIs between Fe-Fe pairs are dominated by RKKY interactions[32,33,34]. Since the RKKY interaction is mediated by conduction electrons, the density of states near $E_F$ plays an important role. To explore this, we calculated the electronic charge density within the energy range from -0.1 eV and 0.1 eV relative to $E_F$, as displayed in the inset in Fig. 2. Obviously, the electronic charge density in this energy range is localized around the kagome layers, indicating that conduction electrons primarily propagate within the kagome layers. Consequently, FeGe and FeSn can be regarded as quasi-2D systems when discussing properties associated with conduction electrons. The corresponding RKKY interaction can thus be described using the recently developed theoretical framework for 2D magnetic materials[42]. In this theory, the exchange energy of 2D RKKY interaction is expressed as

$$J_R(r) = cZ \frac{\cos 2\sqrt{4\pi(1+p)Z/\sqrt{3}d}}{\left(2\sqrt{4\pi(1+p)Z/\sqrt{3}d}\right)^2}, \quad (1)$$

where $c$ is a constant proportional to the density of states at $E_F$, $Z$ is the number of conduction electrons per unit cell, $p$ is the spin polarization at $E_F$, and $d$ is the scaled distance $r/a$. For a small variation of the lattice size, the scaled periodicity of the RKKY interaction is usually independent of the lattice constant. This is a unique feature of 2D magnetic materials for extracting the RKKY interaction from the total MEIs.

Since the RKKY-type oscillation along the <100> direction is much more pronounced than along other directions (Fig. 3), manifesting the anisotropic electronic properties, we focus on $J_{<100>}(r)$ to analyze the RKKY interaction. Figure 4 presents the calculated $J_{<100>}(r)$ for different lattice constants. For FeGe, $J_{<100>}(r)$ at short distances (particularly $J_1$ and $J_3$) varies strongly with lattice constant, illustrating its sensitivity to structural changes. At large distances ($r \geq 2a$), however, $J_{<100>}(r)$ is essentially insensitive to the lattice constant. In contrast, $J_{<100>}(r)$ of FeSn shows less pronounced short-range variations, but the changes remain noticeable. At longer distances beyond $2a$, $J_{<100>}(r)$ exhibits more significant oscillations than in FeGe, with some values still sensitive to the lattice constant.

To extract the RKKY exchange energy, we fitted Eq. (1) using four sets of $J_{<100>}(r)$ values that exhibit the least sensitivity to lattice constant variation. As shown in Fig. 4, $J_1$ and $J_3$ deviate substantially from the RKKY prediction, not only in magnitude but also in sign, because they contain additional contributions except for RKKY interactions. For example, $J_1$ is contributed by both direct MEI and RKKY interaction. The total $J_1$ values are 8.99 meV (FeGe) and 7.92 meV (FeSn), whereas the RKKY contributions are -0.99 meV and -1.35 meV, respectively. This indicates that the RKKY interaction favors AFM coupling for nearest Fe neighbors, acting as a competing mechanism to the intra-layer FM state. The direct exchange energies, obtained by subtracting the RKKY contribution from $J_1$, are 9.98 meV for FeGe and 9.27 meV for FeSn, indicating favorable FM coupling. Apparently, the direct exchange energy is inversely proportional to the Fe-Fe bond length ($d_{Fe-Fe}$) which is 2.48 Å in FeGe and 2.65 Å in FeSn. Considering that other intra-layer exchange energies are relatively minor, $J_1$ is a dominant factor for determining the Néel temperatures of FeGe and FeSn. Since the FM direct MEI is weaker and the AFM RKKY interaction is stronger in FeSn than in FeGe, $T_N$ of FeSn is remarkably lower than that of FeGe.

We next investigated the impact of the CDW phase on the magnetic properties of FeGe. Figure 5 displays the structural feature of the CDW phase. In this phase, a quarter of the $Ge_k$ atoms (notated as $Ge_{k1}$) are displaced out of the kagome plane, and pairs of $Ge_k$ atoms from adjacent kagome layers form Ge-Ge dimers along the *c* axis. This dimerization results in a substantial shortening of the Ge-Ge distance by about 1.3 Å (from 4.05 Å to 2.75 Å), consistent with previous reports[27,28]. The formation of Ge-Ge dimers also induces noticeable displacements of other atoms, although these displacements are much smaller in amplitude. Consequently, both Fe and Ge atoms in kagome layers can be sorted into two types, as indicated in Fig. 5.

In the distorted structure (Fig. 5), the exchange energies between nearest Fe neighbors can be distinguished as $J_{11}$, $J_{22}$, and $J_{12}$, with values of 8.16, 8.35, and 9.24 meV, respectively. The corresponding RKKY contributions are -1.12, -1.22, and -1.36 meV, respectively, yielding direct exchange energies of 9.28, 9.57, and 10.60 meV, respectively. Clearly, the CDW phase affects the direct MEIs more strongly than the RKKY interactions. In addition, all of these values are inversely proportional to the Fe-Fe bond lengths, i.e. 2.49, 2.48 and 2.47 Å for $J_{11}$, $J_{22}$, and $J_{12}$, respectively. However, the small difference in bond length suggests that it is not the primary mechanism driving the notable change in exchange energies. Instead, the symmetry breaking in the CDW phase modifies Fe-Fe hybridization, which may play a major role in altering exchange energies. For the nearest inter-layer Fe neighbors, the exchange energies are $J_{c11}$ = -0.98 meV, $J_{c22}$ = -0.87 meV, $J'_{c11}$ = -1.40 meV, and $J'_{c22}$ = -1.46 meV, all differing from the corresponding values in the ideal structure (Table I). The CDW phase also causes a slight enhancement of the Fe spin moment by about 0.1 $\mu_B$. Overall, these effects strengthen the magnetic state of FeGe. Nonetheless, since its transition temperature (~ 100 K) is far below $T_N$ (~ 410 K), the CDW phase does not influence the performance of FeGe under ambient conditions.

It is known that biaxial strain is an effective approach for tuning the electronic and magnetic properties of 2D materials[43,44]. To explore this, we applied moderate strains ranging from -5% to 5% to both FeGe and FeSn, where negative and positive values denote compressive and tensile strain, respectively. Figure 6 shows the PDOS of Fe, $Ge_k$ and $Sn_k$ atoms under strain. For Fe atoms, two dominant peaks, labeled 'A' and 'B' in Fig. 6(a), correspond to the antibonding and bonding states, respectively. Under compressive strain, the energy separation between the peaks 'A' and 'B' increases relative to the strain-free case, reflecting strengthened Fe–Fe hybridization: shorter bond lengths push the antibonding states upward in energy and the bonding states downward. Conversely, tensile strain reduces this separation. Nevertheless, the overall energy separations between the bonding and antibonding states in FeSn are smaller than those in FeGe, due to the larger lattice constant and correspondingly longer $d_{Fe-Fe}$. The PDOS of $Ge_k$ and $Sn_k$ atoms also exhibit similar strain-induced trend for hybridization with Fe states, as shown in Fig. 6(b) and 6(d). These results demonstrate that biaxial strain strongly influences the electronic structure of FeGe.

Figure 7 displays the local spin moment ($M_S$) of Fe atoms under biaxial strain. Interestingly, $M_S$ of Fe in FeSn depends almost linearly on biaxial strain, increasing from 1.72 $\mu_B$ to 2.27 $\mu_B$ as strain changes from -5% to 5%. In contrast, $M_S$ of Fe in FeGe remains nearly constant between -5% and 2%, but increases linearly from 2% to 5%. These distinct responses can be interpreted by the PDOS of Fe in Fig. 6(a) and 6(c).

For strain-free FeGe, a small but relatively broad peak locates above $E_F$ (approximately 0.2–1.2 eV) in the spin-up channel, as shown by the grey shade area in Fig. 6(a). Under compressive strain which results in shorter Fe-Fe bonds, this peak shifts upward due to stronger Fe-Fe hybridization. However, this shift does not impact the electron occupancies noticeably in the spin-up channel. Meanwhile, the spin-down states near $E_F$ also remains nearly unchanged. Consequently, the electron numbers in both spin-up and spin-down channels, denoted as $n_\uparrow$ and $n_\downarrow$, are preserved against compressive strain, so that $M_S$ ($= n_\uparrow - n_\downarrow$) is unaffected. Under tensile strain which enlarges Fe-Fe bonds, this peak is pushed downwards. However, $M_S$ begins to change only when the peak crosses $E_F$ at tensile strain of ~2% (Fig. 7), owing to the notable change of electron occupancy.

For strain-free FeSn, the corresponding peak lies mostly below $E_F$, approximately ranging from -0.8 to 0.2 eV, so both compressive and tensile strains significantly alter electron occupancies, leading to a continuous change in $M_S$. In fact, the strain-induced change of electron occupancies is more evident in the PDOS of Ge$_k$ and Sn$_k$, as shown in Fig. 6(b) and 6(d). Note that the peak in FeSn has been pushed entirely above $E_F$ under compressive strain of -5% [Fig. 6(b)], suggesting that further compression will not change electron occupancies remarkably, similar to the situation in FeGe. Accordingly, we extended compressive strain to -10% for FeSn. Figure 8(a) obviously shows that $M_S$ of Fe in FeSn indeed remains almost unchanged under compressive strain from -10% to 5%.

Interestingly, a unified trend is revealed when $M_S$ of Fe in both materials is presented as a function of $d_{Fe-Fe}$ [Fig. 8(a)]. For $d_{Fe-Fe} > 2.52$ Å, $M_S$ increases linearly with increasing $d_{Fe-Fe}$, while below this threshold it is almost independent of $d_{Fe-Fe}$. The saturation values are ~1.5 $\mu_B$ for FeGe and ~1.7 $\mu_B$ for FeSn, with the difference attributed to the different electronegativities between Ge and Sn. Since Fe-Fe hybridization dominates in this kagome lattice, it is reasonable that both electronic and magnetic properties are closely associated with $d_{Fe-Fe}$. Therefore, the $M_S$-$d_{Fe-Fe}$ relation revealed in Fig. 8(a) should be considered as a general feature of kagome magnetic materials. Nevertheless, because experimentally feasible strains are typically within ±5%, our subsequent discussion of MEIs is restricted to this range.

The calculated $J_1$ under strain is plotted in Fig. 8(b). In both FeGe and FeSn, $J_1$ decreases monotonically with increasing $d_{Fe-Fe}$ (or equivalently strains). Remarkably, when $J_1$ values of FeGe and FeSn are considered together, $J_1$ follows an approximately linear dependence on $d_{Fe-Fe}$. Hence, we fitted the $J_1$ values with a linear function: $J_1 = \alpha\, d_{Fe-Fe} + \beta$, as presented by the dashed blue line in Fig. 8(b). Obviously, the fitted curve covers most calculated $J_1$ values. It is reasonable because $J_1$ is dominated by direct MEI which is approximately proportional to $d_{Fe-Fe}$. Additional Fe-Ge and Fe-Sn hybridizations cause deviations near the crossover region due to the distinct different electronegativities of Ge and Sn.

Although $J_1$ shows a simple monotonic dependence on biaxial strain, other $J(\vec{r})$ values are more complex, as seen in Fig. 9(a). For instance, $J_{c1}$ first increases as biaxial strain increases, then decreases for strain larger 3%. $J_2$ of FeGe decreases monotonically with increasing biaxial strain, but $J_2$ of FeSn exhibits oscillatory character. Using all $J(\vec{r})$ values, we employed the Heisenberg model and performed Monte Carlo simulations to obtain Néel temperatures of FeGe and FeSn. Without strain, the calculated $T_N$ is 420 K for FeGe and 380 K for FeSn, agreeing with the experimental measurements very well. Under biaxial strain, the strain dependence of $T_N$ is non-monotonic for both FeGe and FeSn, as shown in Fig. 9(b), indicating that indirect MEIs are important to determine $T_N$, although $J_1$ is the dominant factor. For FeGe, $T_N$ decreases monotonically for strain varying from -5% to 2%, then increases at larger tensile strain. Intriguingly, $T_N$ of FeGe can be enhanced to 540 K under compressive strain of -5%, while under tensile strain $T_N$ still remains above 350 K. For FeSn, compressive strain also strongly enhances $T_N$, up to 450 K at compressive strain of -4%, whereas tensile strain makes $T_N$ fluctuate around 380 K. Therefore, while the AFM states in both FeGe and FeSn are robust against moderate tensile strain, they can be substantially strengthened by moderate compressive strain, providing a practical route to engineer AFM orders in kagome magnets.

## IV. Conclusions

In summary, we systematically investigated the magnetic properties of kagome antiferromagnets FeGe and FeSn, based on first-principles calculations. We found that the electronic states near $E_F$ are localized within kagome layers, implying that FeGe and FeSn can be regarded as quasi-2D systems for exploring their conduction-electron-related electronic and magnetic properties. The calculated exchange energies exhibit strong anisotropy, with clear RKKY oscillations along the Fe-Fe chains. In both materials, the nearest neighboring exchange energy $J_1$ is dominant, arising from contributions of both direct MEIs and RKKY interactions. The direct MEIs favor FM coupling, while the RKKY interactions prefer AFM coupling. Compared with FeSn, FeGe has stronger direct MEIs but weaker RKKY interactions, resulting in a substantially higher Néel temperature $T_N$. In addition, the CDW phase of FeGe further enhances its magnetic state below the transition temperature. Interestingly, the local spin moment $M_S$ of Fe atoms in both materials follows a unified dependence on the Fe-

Fe bond lengths: $M_S$ increases linearly with increasing $d_{Fe-Fe}$ when $d_{Fe-Fe} > 2.52$ Å, while it retains almost constant at shorter distances. Similarly, $J_1$ scales approximately linearly with $d_{Fe-Fe}$, with shorter $d_{Fe-Fe}$ yielding larger $J_1$. In contrast, other exchange energies display more complex variations with $d_{Fe-Fe}$ (or equivalently, biaxial strain), leading to non-monotonic strain dependence of $T_N$. Nevertheless, moderate compressive strain can significantly enhance $T_N$ in both FeGe and FeSn, offering an effective pathway for engineering magnetic orders in kagome magnets.

**Acknowledgements**

This work is supported by the Program for Science and Technology Innovation Team in Zhejiang (Grant No. 2021R01004), the Six Talent Peaks Project of Jiangsu Province (2019-XCL-081), the start-up funding of Ningbo University and Yongjiang Recruitment Project (432200942).

Table I. Exchange energies (in meV) in FeGe and FeSn, as labeled in Fig. 1.

|      | $J_1$ | $J_2$ | $J_3$ | $J'_3$ | $J_{c1}$ | $J_{c2}$ |
|------|-------|-------|-------|--------|----------|----------|
| FeGe | 8.99  | -0.76 | -0.27 | 0.53   | -1.25    | 0.01     |
| FeSn | 7.92  | -0.87 | -0.43 | 0.59   | -1.12    | 0.21     |

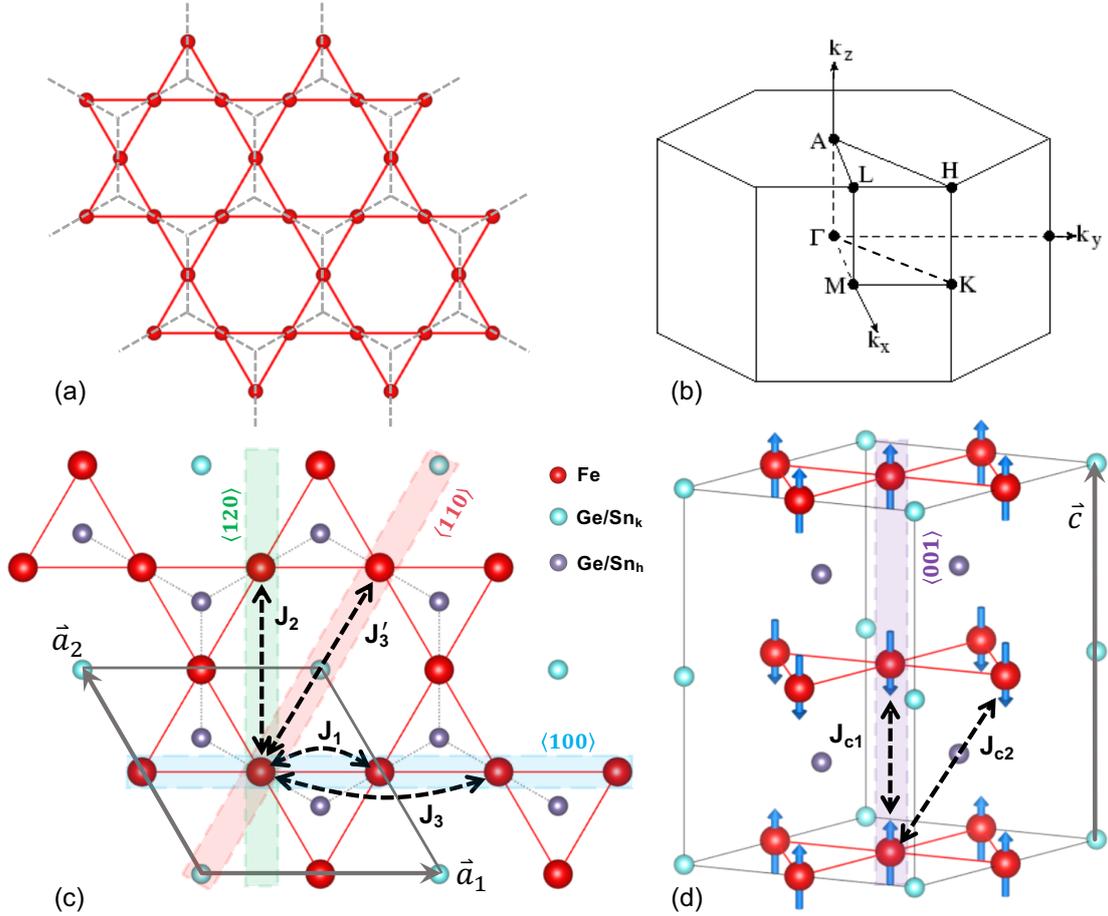

Fig. 1. (a) Top view of a two-dimensional kagome lattice. The dashed network outlines the maternal honeycomb lattice. (b) Brillouin zone of a three-dimensional hexagonal lattice, with the special k points indicated. (c, d) Top and side views of the crystal structure of FeGe and FeSn. Red and cyan spheres represent Fe and Ge/Sn atoms in the kagome layers, respectively, while grey spheres stand for Ge/Sn atoms in the honeycomb layers. The grey arrows indicate the lattice vectors. Shadow areas with different colors highlight specific directions used to extract exchange energies. $J_i$ and $J_{ci}$ represent intra- and inter-layer exchange energies, respectively, for Fe-Fe pairs connected by the corresponding black double-headed arrows. Blue arrows centered on Fe atoms symbolize their spin moments.

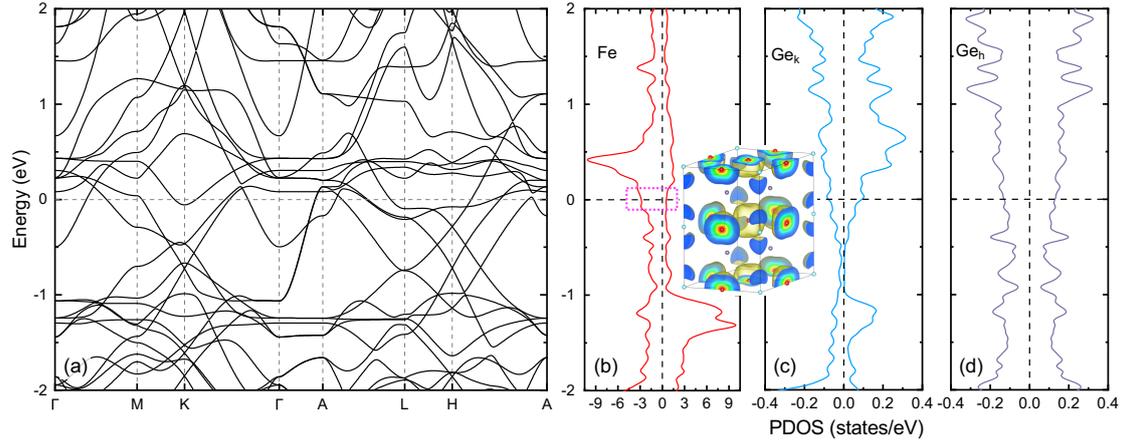

Fig. 2. Electronic properties of FeGe. (a) Band structure. (b–d) Projected density of states (PDOS) of Fe, Ge$_k$ (in kagome layer) and Ge$_h$ (in honeycomb layer). The Fermi level is set to zero energy. The inset crossing (b) and (c) depicts the electronic charge density within the energy window from -0.1 eV and 0.1 eV as marked by the pink dashed rectangle in (b).

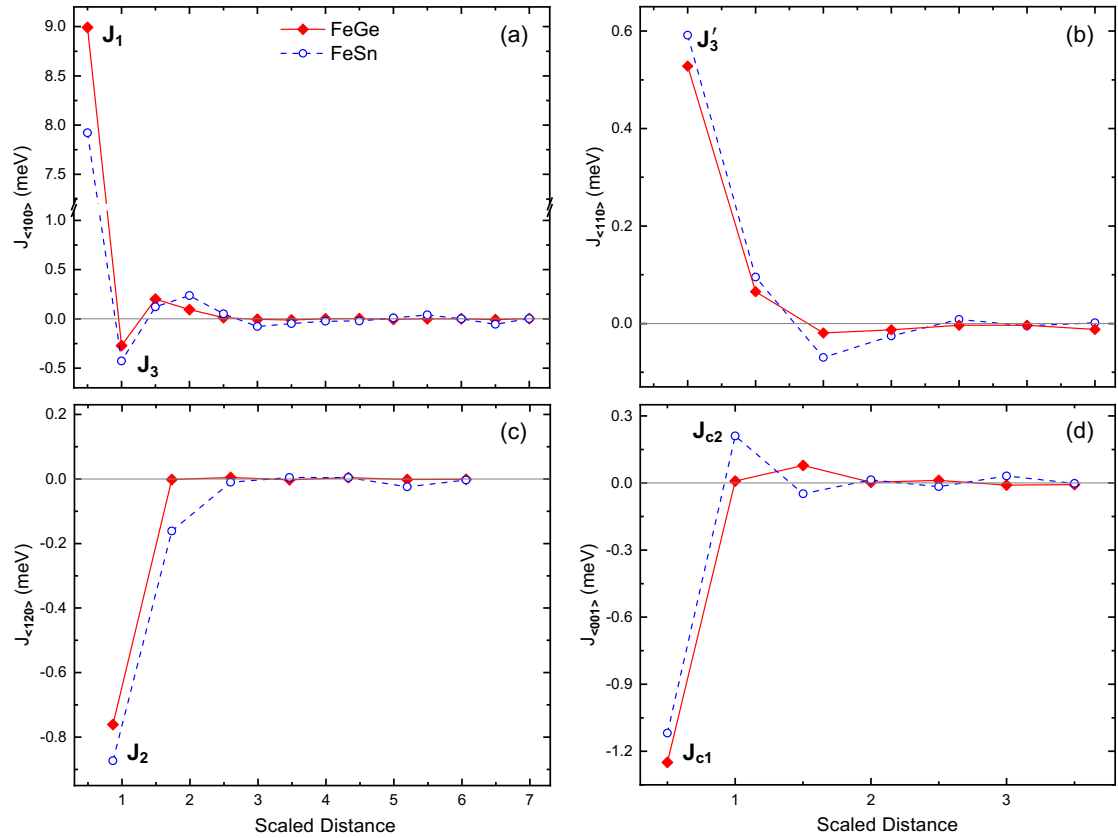

Fig. 3. Exchange energy $J(\vec{r})$ as a function of scaled distance along the selected directions shown in Fig. 1. The scaled distance is defined as $r/a$ in (a–c) and $r/c$ in (d). Here, $r$ is the Fe-Fe distance, whereas $a$ and $c$ are the in-plane and out-of-plane lattice constants, respectively.

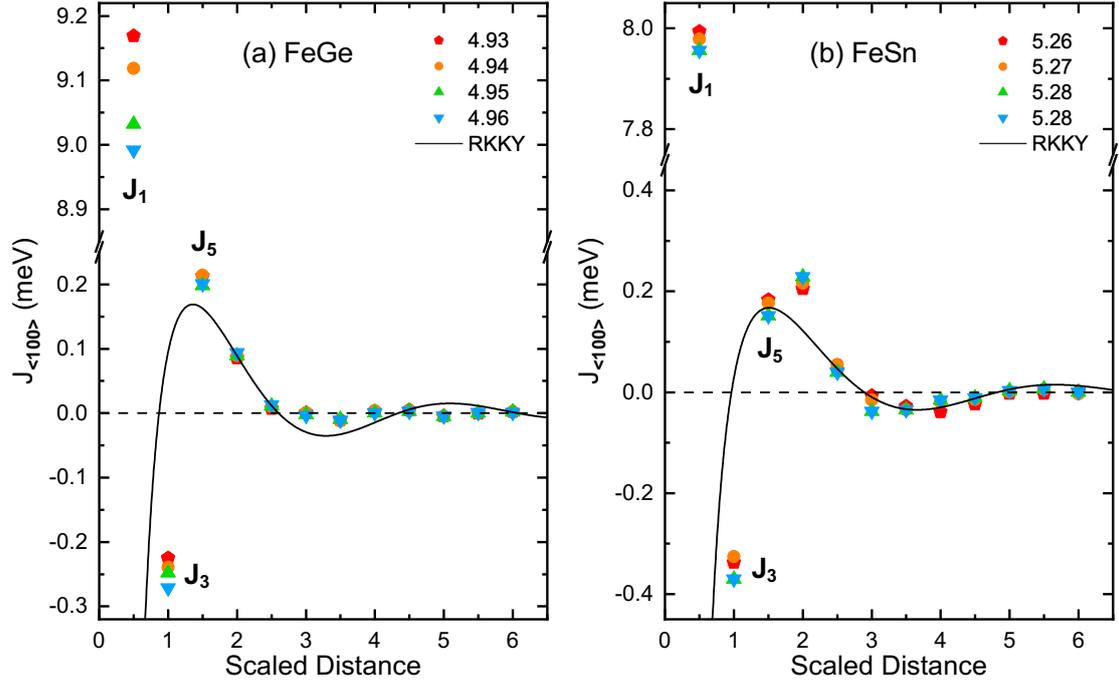

Fig. 4. Exchange energy $J_{<100>}(r)$ as a function of scaled distance for four different in-plane lattice constants. Symbols stand for values obtained from first-principles calculations, while the solid curve represents the RKKY exchange energy fitted by Eq. (1).

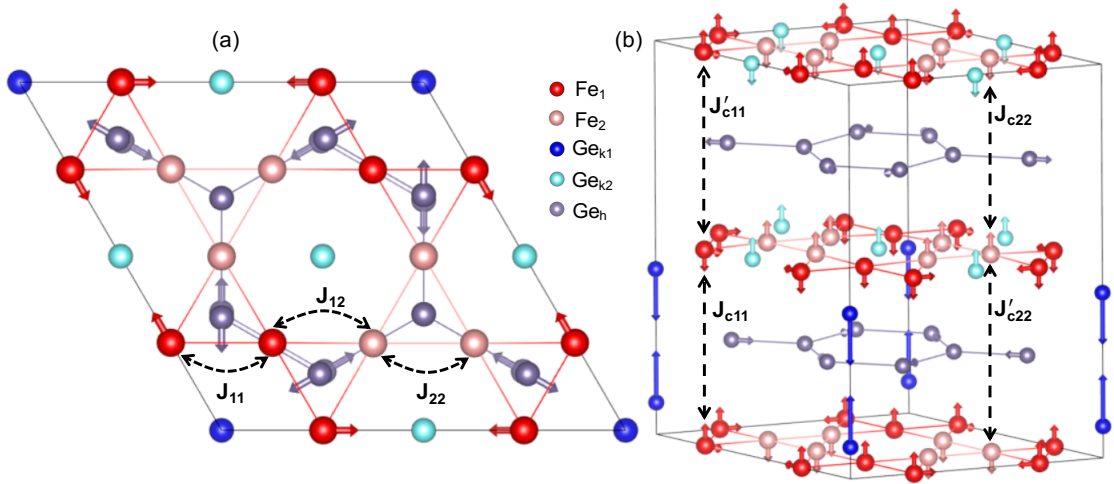

Fig. 5. (a) Top and (b) side views of atomic displacements of the charge density wave phase of FeGe. Arrows centered on atoms indicate the directions and magnitudes of their displacements. $J_{ij}$ and $J_{cij}$ denote intra- and inter-layer exchange energies, respectively, for nearest neighboring Fe-Fe pairs connected by black double-headed arrows.

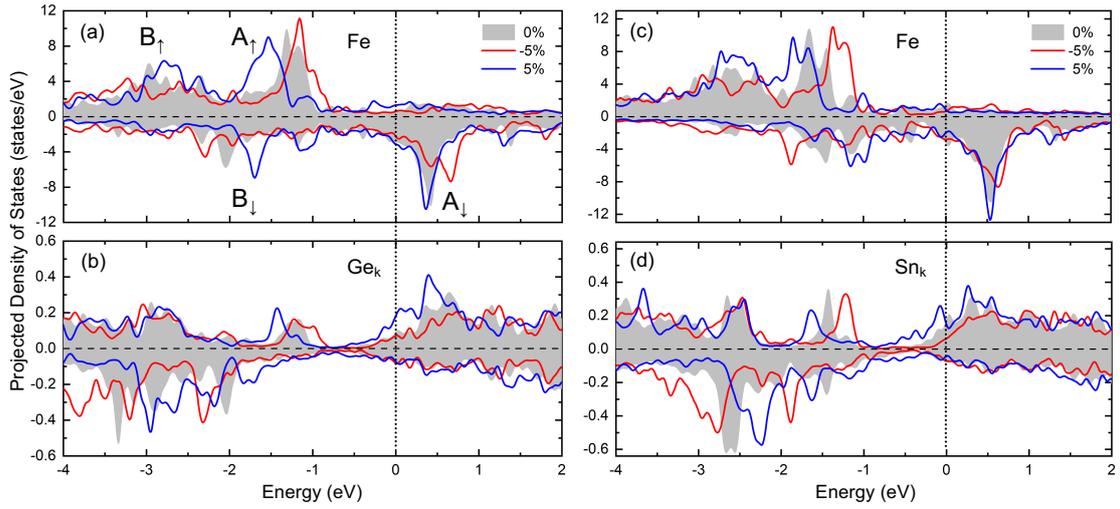

Fig. 6. Projected density of states of Fe and $Ge_k/Sn_k$ (in kagome layer) with biaxial strains, compared with the strain-free case (grey shadow areas). (a, b) for FeGe, and (c, d) for FeSn. The Fermi level is set to zero energy. 'A' and 'B' in (a) notate the antibonding and bonding states, respectively, originating from Fe-Fe hybridization within the kagome layer. The subscript arrows indicate spin-up and spin-down channels.

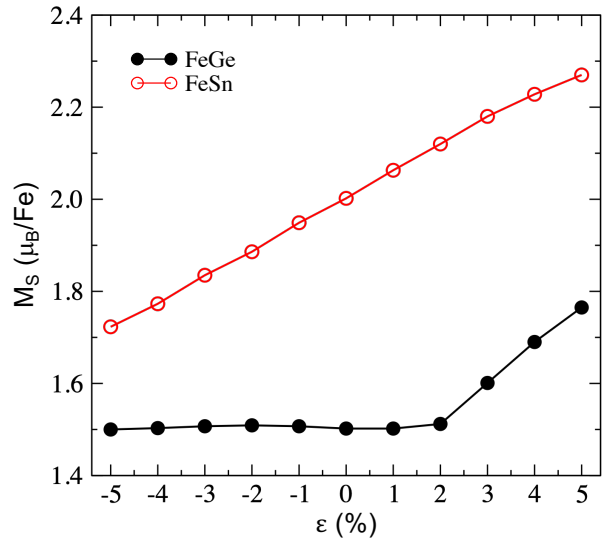

Fig. 7. Spin moment ($M_S$) of Fe as a function of biaxial strain ($\varepsilon$).

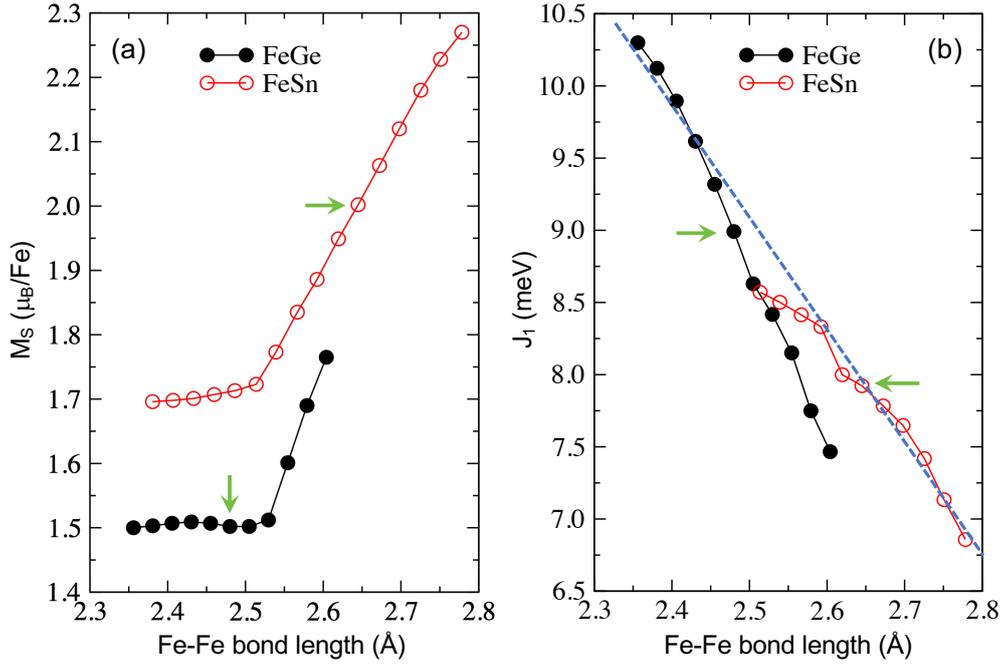

Fig. 8. (a) Spin moment ($M_S$) of Fe and (b) nearest neighboring exchange energy ($J_1$) as functions of the Fe-Fe bond length ($d_{Fe-Fe}$). The corresponding biaxial strain is from -5% to 5% for FeGe in both (a) and (b), while from -10% to 5% for FeSn in (a), and from -5% to 5% for FeSn in (b). Green arrows designate the values at the strain-free lattice constants of FeGe and FeSn. The blue dashed line in (b) represents a linear fit given by $J_1 = \alpha\, d_{Fe-Fe} + \beta$, with $\alpha$ = 7.75 (meV/Å) and $\beta$ = 28.46 (meV).

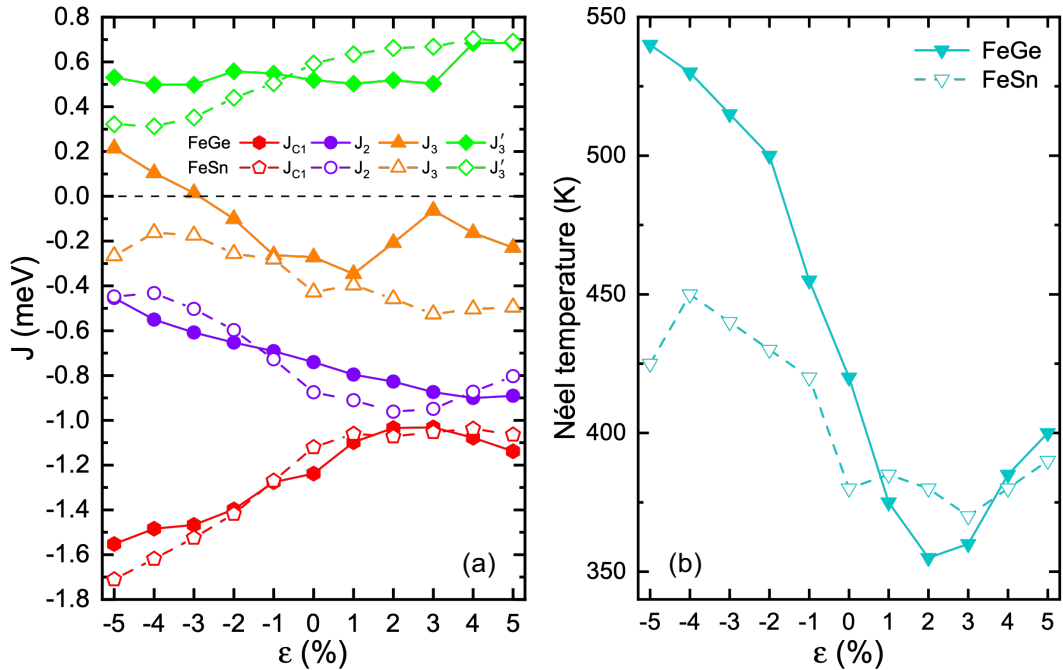

Fig. 9. (a) Exchange energies and (b) Néel temperatures of FeGe and FeSn as functions of biaxial strain ($\varepsilon$).